\documentclass[12pt]{article}
\usepackage{latexsym}
\usepackage{graphicx}
\usepackage{amsmath}
\usepackage{hyperref}
\usepackage{amssymb}

\begin{document}

\begin{center}
\textbf{\Large Model equations and structures formation for the media with memory} \vspace{0.5 cm}

 O. Makarenko$ ^{\dagger,}$\footnote{e-mail:
\url{makalex51@gmail.com}}, \vspace{0.5 cm}

$ ^{\dagger}$Institute for Applied System Analysis NTU ''KPI''

Politekhnichna st., 14, Kyiv, Ukraine, 03056
\end{center}

\begin{quote} \textbf{Abstract. }{\small 
We propose new types of models of the appearance of small- and
large-scale structures in media with memory, including a hyperbolic modification of the Navier–Stokes equations and a class of
dynamical low-dimensional models with memory effects. On the
basis of computer modeling, the formation of the small-scale structures and collapses and the appearance of new chaotic solutions
are demonstrated. Possibilities of the application of some proposed
models to the description of the burst-type processes and collapses
on the Sun are discussed.
}
\end{quote}


\vspace{0.5 cm}

\section{Introduction}

\noindent The basic equations for heat transfer and hydrodynamics are
usually parabolic heat equations and the Navier-Stokes
hydrodynamic equations. But these equations lose their applicability in extended
media, when characteristic scales of parameters change less than
correlation time and correlation length (relaxation or memory and
nonlocality effects). Many examples of non-applicability were
found in turbulence. Therefore, more correct equations should be
applied in such cases. There are some well-established facts in
the theoretical physics on the description of transport processes.

The first famous idea is the existence of hierarchy of description
levels. If there are $N \gg 1$ particles we have many levels for
description: $N$ deterministic dynamical laws of Newton for
particle movement; Liouville equation for $N$--particle
distribution function, Boltzman equation for one-particle
distribution function, hydrodynamic equations for macroparameters
(Naveir-Stokes class equations), thermodynamic equations. The
choice of the level of description depends on the measure of
deviation from equilibrium.

The second important idea is the existence of many interrelating
relaxation processes and many time and space scales with different
relaxation times and lengths. The memory and nonlocality effects
are common for all levels. Each level of description has its own
specific type of chaos, autowave solutions, collapses and so
on. It should be stressed that each level of description has its
own model equations with typical behaviour of the solution. The
problem of defining typical chaotical behaviour for a given
level is especially interesting.

The turbulence is a bright example of such complex phenomena. It
is well known,  turbulence is encountered  very often in broad
classes of natural processes. It is widely recognized that
turbulence is a strongly non-equilibrium phenomenon. Investigators
dealing with it in the specific branches of science, as a rule,
formulate what is turbulence on the intuitive level. This results
in the lack of the generally accepted and unified definition of
turbulence. In concrete cases one can encounter different types of
turbulence. This difference is fixed, for example, in the plasma
theory (strong and weak turbulence).

Till now,  strict and evident in all cases,  definitions of
turbulence  are absent. To imagine a set of possibilities   in
this question we will recall  a few descriptions of turbulence
mainly related to hydrodynamics.

"Turbulent motion of liquid at  large enough  values of the
Reynolds number is characterized by the extraordinarily irregular,
disorderly change of speed in  time in every point of stream (the
developed turbulence). The velocity  always  pulsates near the
some mean value" \cite{vladimirov:dau}

"Turbulent motion is, generally speaking, vortical"
\cite{vladimirov:dau}.

"So, the turbulence is the vortical flow of  viscous liquid which
is stochastic in the sense that it is characterized by:  1)
sensitive dependence on initial conditions (this caused by
exponential divergence initially closed phase trajectories); 2)
that all phase trajectories are dense almost everywhere, thus, any
initial nonequilibrium distributions of probability in phase space
tend  to the limiting equilibrium distribution; 3) mixing in phase
space (and, as a result, ergodicity, by the rapid fading of
temporal correlation functions and continuity of frequency
spectrums). The developed turbulence possesses additionally 4)
multimodes, and, as a consequence, manifests the  chaos in its
spatial structure at any fixed moment of time" \cite{monin}.

"This motion is so difficult and so not  studied enough, that even
the question: what is turbulence? is difficult to answer. What
kind of  basic features  of turbulent motion are there?"
\cite{klimontovich}.

"One of the most considerable features of turbulence is
co-operation of order and accident... An accident in turbulence is
related to the extreme sensitiveness to disturbances. At the same
time, the chosen statistics are roughly stable at disturbances"
\cite{Kraichnan}.

"What is  turbulence? Some light for this  difficult purpose is
explained by the study of offensive  turbulence or weak turbulence
which is an aspect of  chaos" \cite{Ruelle}.

"The following definitions of turbulence can be offered... A
turbulent flow should  be unpredictable in the sense that a small
uncertainty in initial time will grow so that the strict
deterministic prediction of its evolution is not possible, the
second, a turbulent flow should satisfy the property of the
increasing mixing" \cite{Lesier}.

Still, a number of determinations is presented by the theory of
plasma turbulence that, as will be argued below, it has definite
interest and is hydrodynamic turbulence, so, the most common
descriptive determination: "Any state of plasma with the strongly
developed noises and vibrations is accepted as turbulent"
\cite{kadomtsev}.

The selection in plasma turbulence of collective degrees of
freedom and construction of turbulence on their basis is
important.

"Presentation which is usually inlaid in notion of turbulence
substantially differs from the notion of molecular motion.
Speaking, for example, about motion of any element of liquid,
motion of some macroscopic volume is implied, containing the large
number of molecules... In other words, hydrodynamic motion of
liquid corresponds to the definite collective degrees of
freedom... It is possible  to define the concept of "turbulence":
the turbulent motion of macroscopic bodies is such a motion in
which the collective degrees of freedom are intensively excited
and carried an accidental character" \cite{tsytov1}.

As a rule,  it is assumed  that turbulence in plasma and liquid
are different. In \cite{tsytov}  the  comparison of turbulence in
plasma  with turbulence of incompressible liquid is given. It
turns out that  instability of liquid causes the excitation of
whirlwinds and there are difficulties with the flows of energy. It
is considered that the indicated difficulties have  a principle
character and are connected with the strong turbulence
 in incompressible liquid. This is expressed by the following:  whirlwinds
generally do not have their  own frequencies and the time of
transmission of energy from one whirlwind to a neighbouring one is
about one turn of the whirlwind.

Unlike the motion in incompressible liquids,  in plasma there are
a lot of collective oscillating motions which have their  own
definite frequencies. Time of transmission of energy  of these
vibrations to the neighbouring scales (or neighbouring wave
numbers) can essentially exceed the characteristic period of
vibrations $\omega_k^{-1}$. Because of this "elasticity" of
collective motions a small parameter $\omega=\tau\omega_k^{-1} \ll
1$ can appear. It is considered that the presence of this small
parameter allows to use the regular methods of decomposition on
energies of turbulence and build the theory of weak turbulence
\cite{tsytov}. The role of dispersion for plasma turbulence is
 underlined  here.

So, in \cite{shapiro} it is indicated that in  incompressible
liquids  the dispersion is absent. Therefore, in this situation we
see the turbulence of whirlwinds and transmission of energy
between them. Plasma possesses many excitations with  dispersion,
co-operation, and weak turbulence \cite{shapiro}. In addition, in
plasma there is strong turbulence of solitons. Let us note here
that the results of investigations presented in the cited work
presumably shows that   a sharp barrier between the types of
turbulence in plasma and liquid is not present and the situation
turns out to be more difficult and interesting. Plenty of
determinations indicates that common determination of turbulence
is absent and in every concrete process there can be one specific
type of turbulence. This distinction is fixed, for example, in the
theory of plasma (strong and weak turbulence).

But even if  a rigorous definition exists, we must use some
mathematical idealization in the theoretical study of this
complicated phenomenon. For distributed objects we usually take
the nonlinear PDEs or integro-differential equations. Then
mathematical models are investigated on stochasticity connected
with the unstability in systems.

The mathematical object on the hydrodynamic level is the system of
partial differential equations for velocity, pressure and
temperature. Such a system includes equations of motion as well as
the constitutive equations, which give the expressions for the
deformations versus strain dependence \cite{Batchelor}. Instant
connections between displacements and strain lead to, in the case
of the incompressible fluid, the Navier-Stokes  equations. The
memory effects take on the form of integral constitutive
equations. The simplest exponential kernels lead to the
constitutive equations introduced formerly for visco-elastic
media. In such cases we obtain the simplest modification of the
Navier-Stokes equations, namely,  the so-called equations for the
Maxwell media (\cite{oskol, osk, mak_swir2005}).

Such modified system \cite{oskol, osk, mak_swir2005} is still a very complex mathematical
object. In order to understand its properties, one can investigate
 the simpler model equations. It is well known that such a model
equation for the NavierStokes system is the one-dimensional in
space nonlinear Burgers equation \cite{tsytov}. This equation was
introduced in 1948 empirically, and later derived within the
asymptotic approach. It has been recently proposed  \cite{shapiro}
to include a second-order derivative with respect to time in the
Burgers equation. The main goal of such a modification was the
incorporation of memory effects. Then, we have the one-dimensional
in space model equation, which is the so-called hyperbolic
modification of the Burgers equation \cite{monogr,Dan_Kor_Mak}.

Its numerical and analytical studies reveal many curious
properties. One of the most interesting properties is the
existence of blow-up (collapse) solutions. However, not long ago,
the two and three-dimensional model equations for visco-elastic system
did not exist. Following the lines of deriving
the Burgers equation we put forward  multi-dimensional model
equations. Analysis of the full system \cite{oskol, osk, mak_swir2005} and
physical properties of systems with memory (fast heat processes,
plasma, turbulence, statistical mechanics, visco-elasticity)
allows to pick out some necessary properties that should be
accounted for by model equations. Such properties are viscosity,
mass transfer by convection and finite speed of disturbances
propagation. Besides, the properties of such  model equations
should resemble in some specific cases the behaviour of original equations
\cite{oskol, osk, mak_swir2005} and, what is more important, the behaviour of
real objects (for example, the stability properties). Our analysis
leads to some new model equations with such properties. The
simplest two- and three-dimensional equations have the form
$$
\mu \frac{\partial ^2 \vec u}{\partial t^2 } + \frac{\partial \vec
u}{\partial t} + u_k\frac{\partial \vec u} {\partial x_k} = c
\triangle \vec u.
$$
Also, the nonlocality may be incorporated in the model equations.

\section { The choice of model equations}%

Since a deep theoretical description of  mathematical models for
hydrodynamic processes can be carried out by means of the methods
of theoretical physics, we very briefly describe some main
concepts from statistical physics relevant to modelling equations
choice.

Begining from the works of N.Bogoliubov, M.Born, M.Green, J.
Kirkwood and J. Ivon, the canonical approach in the theoretical
physics is as follows. Let us consider the medium constituted from
$N$ independent particles. Then, in classical physics we can
describe the particle movement precisely by a system of ordinary
differential equations (the Newton equations). But the statistical
physics consider the ensemble of the system by introducing
distribution function $f_N(x_1, x_2,... , x_N; t)$ for particles
distribution probability at time $t$. The function $f_N$ is
evaluated from the Liouville equation
\begin{equation}\label{mak:1}
    \frac{\partial f_N}{\partial t}+\left\{f_N,\,H\right\}=0,
\end{equation}
where $H$  is the Hamiltonian of the system and $\{\cdot\}$ are
the Poisson brackets.

However, the function $f_N$ is too informative for the
hydrodynamic phenomena description. Usually all necessary
parameters are macroparameters (for example temperature, pressure,
and velocity: $T$, $P$ and $V$). The main leading principle in
such a case is the reduction of the description parameters set.
The reduction procedure deals with some hierarchical levels.
Firstly, by integrating on some variables in phase space, we can
go to $f_1-$one-particle distribution function with the BBGKI
chain of equations for $s$-particle distribution functions. Note
that for $f_1$ we can receive the well known Boltzman equation.
These stages with distribution functions are named kinetical.

Further averaging with one-particle distribution function leads to
ma\-cro\-para\-meters $T$, $U$, $P$. Usual procedures lead to well
known equations of the hy\-dro\-dyna\-mic type: the parabolic heat
equation, the Navier-Stokes equations and so on. But a more
correct description leads to more difficult equations with memory
effects. The reason of memory effects origin under reduction
processes is very well described in theoretical physics since the
works of H. Mory, R. Zwanzig, R. Picirelly, D.  Zubarev and many
others, see review in \cite{res,Zub_Tish,zub92,Makar_97}. In this
approach we receive hydrodynamic equations for macroparameters
with some constitution equations relating to ma\-cro\-vari\-ables.
In general, such constitution equations have the form of
integro-differential equations \cite{monogr}.

Let us make some comments concerning the reduced description.
Reduction of description takes place not only at the level of the
BBGKI chain. Thus, even having equations of hydrodynamics and
solving them by approximate methods, we get the chain of reduced
descriptions corresponding to different accuracy of methods. This
easy to see  in the  Galerkins method for the Navier-Stokes
equation \cite{temam2}. It is known that  the construction of
solutions in the form of a series in terms of trigonometric
functions leads to the system of three ordinary differential
equations (the Lorenz system) with the chaotic behaviour. If we
take into account a higher  number of modes $M>3$, then other
types of chaotic or even periodic behaviour are observed at
$M>M^{\prime\prime}$.

It is clear that different Galerkins systems derived at different
numbers $M$ correspond to different levels of reduced description
(with $M$ leading parameters). This can help to understand the
sense of such chaotic behaviour and  answer the
 question: can the low dimensional dynamical systems
 transfer the chaotic behaviour of the initial system of PDE.

 Here, we can reason in the same way as in  the case of the
Liuvills equations. If we had the complete phase portrait of the
infinite dimensional Galerkin system of ODE, we would pass the
properties of a system of PDE completely and would know the
trajectories exactly.

In order to  substantiate the low dimensional dynamical system and
the limiting transition as $M\rightarrow \infty$ there is a way
appearing from the classical theory of difference schemes when
convergence and accuracy are proven as $M\rightarrow \infty$.

However, it is  very difficult  and in addition, presumably for
any numerical schemes (at fixed $M$), a system with such a
difficult phase portrait can be found, that  a numerical scheme
will not be able to correctly  pass the behaviour of a solution.
However,  in many cases we do not need such detailed information
about the solutions, and it is important to know only some general
patterns. Here, the situation is the same as in the description of
$N$ particles  a deterministic way or by means of the distribution
function.

Therefore, the applicability of the finite-dimensional dynamical
systems as $M\rightarrow\infty$ should be considered  in the sense
of characteristics of chaos by the bringing of probabilistic
conceptions (for example, the limit of invariant measures). In
addition, it is well known that the application of probabilistic
methods in the theory of cellular automatons makes the proof
easier \cite{Presutti}.

Probabilistic considerations also allow to put forward a question
about the correct transmission of chaotic behaviour by means of
different finite-dimensional projections. Similar is done for the
infinite - dimensional Bogolyubovs chains in the so-called
lattice systems \cite{Vidybida}.

We should especially note  the reasons of memory appearance and
the nature of memory in the given context. From Maxwell's
considerations of the model example, the memory appears because of
 accounting the delay. In the thermodynamics of
media with memory,  the memory is simply postulated in
integro-differential relations for the flux of heat and the tenzor
of stress with the defined kernels of integral relations. But the
exact type of relations and concrete expressions for kernels can
be derived  in statistical physics by means of reduced
description, when information about the "unimportant" degrees of
freedom are taken into account by implicitly  through kernels. The
essence of this phenomenon was found out in the so-called
projection method of Mori  Zvantzig's type \cite{Zwanzig}, and
in the method of non-equilibrium statistical operator by
D.~Zubarev \cite{zub,zub55}.

In this approach, the function of distribution $f$ is divided into
 two parts: $f=f_1+f_2$. For example, as a rule,  $f_1$ is chosen
 as the locally-maxwellian distribution depending only on the relatively
slowly changing hydrodynamic parameters $T$, $p$, $u$. Part $f_2$
carries information about quickly changing micro-motions. Then,
the application of  $f_1$ leads to hydrodynamic equations which
should contain contributions of the  "rapid" part $f_2$.

In order to do this,  the operator of projection $P$ is introduced
so that $f_1=Pf$, $f_2=(1-P)f$. Then, one can use the solution
 of the Liuvills equation in the form  $f(t)=\exp \left(
-itL\right) f(0)$, where $L$ is the Liuvill's operator connected
with the  Hamiltonian  system. After some manipulations
\cite{Zwanzig,pic,zub55} the expression for $f_2$ is obtained
\begin{equation}\label{eqn2_8}
f_2=e^{ -i(t-t_0)(1-P)} f_2(t_0)
-i\int_0^{t-t_0}dt^\prime e^{
-it^\prime (1-P)L}(1-P)Lf_1(t-t^\prime)
\end{equation}
and equation for $f_1$
\begin{equation}\label{eqn2_9}
    \frac{\partial f_1(t)}{\partial t}=-iPLf_1(t)-iPL e^{
 -i(t-t_0)(1-P)}f_2(t_0)
-\int_0^{t-t_0}dt^\prime e^{
-it^\prime (1-P)L}(1-P)Lf_1(t-t^\prime).
\end{equation}
It easy to see that (\ref{eqn2_8}) and (\ref{eqn2_9}) have a
temporal nonlocal character. This is caused  by the division of
function of distribution to pieces. The difficulties with the
choice of the projection operator are well known, however,  the
form of which is not known. They were  overcome in the method of
nonequilibrium statistical Zubarev's operator. According to this
method, the function of distribution is assumed to depend on a
certain set of parameters  of reduced description $a_1$, $...$,
$a_k$, $f=f(a_1,..., a_k)$, where $\{a_j\}$ are certain averages
of the  distribution  function $f$ \cite{zub55}. Note that, in
this case, in principle, we can obtain the same relations for $f$
taking  memory into account.

Concerning the hydrodynamic level of description, functions of
distribution lead to the equations of hydrodynamics in the form of
balance laws for momentum and energy
\begin{equation*} \frac{\partial \rho}{\partial t}+\frac{\partial \rho V}{\partial
x}=0,\quad
\frac{\partial V}{\partial t}+V_k \frac{\partial V_k}{\partial
x}+\mbox{grad } p=\mbox{div } \sigma +F,\quad \frac{\partial
E}{\partial t}+\frac{\partial I_E}{\partial x}=0,
\end{equation*}
where $\rho$ is the density, $V$ is the velocity, $p$ is the
press, $\sigma$ is the stress, $F$ is the thermal force, $E$ is
the energy, $I_E$ is the energy flux.

For closing these equations, integral relations for the heat
fluxes and tenzors of stress are used. A short and clear
presentation of these issues  is given in the appendix to the book
written by Day \cite{day}.

Let us write  these relations for the one-dimensional case
\begin{equation}\label{eq11_12}
\sigma(x,t)=\int_{-\infty}^tdt^\prime \int dx^\prime K_1
\frac{\partial u}{\partial x}(x',t'),\,
q(x,t)
=\int_{-\infty}^tdt^\prime \int dx^\prime K_2 \frac{\partial
T}{\partial x}(x',t').
\end{equation}
In these relations, the  kernels $K_1$ and $K_2$,  derived by
means of uniform procedure, are connected with the correlators of
variables. The correlators can be  calculated with the help of the
function of distribution. For example, for $K_1$ we have
\begin{equation}\label{eqn2_13}
K_1=\left(q(x),q_1(x',-t)\right).
\end{equation}
Here, in (\ref{eqn2_13}) there is an assembly average.

Thus, summing up known results on statistical physics it appears
that the application of a reduced description at a hydrodynamical
level leads to  conservation laws, which are closed by some
integral relations. These integral equations can have a very
difficult form including nonlinear dependence on hydrodynamic
variables. The concrete type of kernels depends on a set of
degrees of freedom involved in the process (this indicates the
necessity of their incorporation  at the reduced description).

In turn, these "necessary" degrees of freedom appear in dependence
of the displacement  from  equilibrium. In equilibrium
thermodynamics are enough to characterize the temperature of a
body as whole. It is easy to see that without consideration of
prehistory effects the  ordinary relations for heat conductivity
in the form of the Fourier law $q=-\lambda
\partial T / \partial x$   and the relations for stress of type
$\sigma =-\nu
\partial u / \partial x$ are derived. These relations lead to the parabolic
equation of heat conduction and to the Navier-Stokes equations
accordingly. The exponential kernels of type $\exp(-t/\tau)$ lead
to the hyperbolic equation of heat conductivity and to the
equations of relaxation hydrodynamics.

Exponential kernels correspond to more exact consideration of
property of remoteness from equilibrium. For an adequate but
 rough description of the processes far from equilibrium, the
hydrodynamical equations may be enough in the beginning  (it is
possible with  nonlinear constitutive relations). But after that,
a hydrodynamic level may be not valid, then  we need the whole
one-particle function $f_1$ (or even multi-particle functions of
distribution) as  parameters of reduced description  (see
\cite{zub59}).

Now, let us apply the aforesaid to the problem of turbulence.
Assume that each level of reduced description has the
corresponding characteristic chaotic behaviour, which may be
called the turbulence corresponding to the given level of reduced
description.  In an implicit form, this has been done in
\cite{kulm85}, and in an explicit form  in \cite{mak1988}. A
similar concept is contained in \cite{klimontovich2}. This
assertion  seems to allow to reconcile the seemingly
irreconcilable points of view on the so-called hydrodynamic
turbulence. Of course, this is a complex phenomenon which can
demand the complete description with the help of function $f_N$
(for example, for the developed isotropic turbulence). However,
the levels of reduced description may be enough for weaker
turbulence, and especially for the study of the initial stages of
its development. Note that at  different levels of description,
the features of a general picture may be different. That is why,
despite
 the disadvantages of the  Navier-Stokes
equations, these equations nevertheless may contain information
about the turbulent behaviour. For example, in \cite{Kraichnan}
the expression of Navier-Stocks turbulence appeared even. It is
possible that the Navier-Stocks equations can pass the initial
stages of loss of laminar flows stability.

We observe other types of turbulent behaviour using the levels of
reduced description with the help of the complete Liuvill's
equation. Thus, the place of many works concerning the
incorporation of  different fluctuations is cleared up.
Small-scale fluctuations can be incorporated  at the level of the
Boltsmans equation in the  integral term (\cite{klimontovich2,lib}
and others). Incorporation of large-scale fluctuations, which
correspond to the collective motions like a transfer of liquid,
are much more difficult.  In this case the Reynolds's procedure of
the separation on the average flow and pulsation components does
not help  because it is based on the system of the Navier - Stokes
equations. Therefore, since the 1960's  plenty of work was devoted
to the considerations of large-scale fluctuations. These works
contain the articles considering the rejection of the hypothesis
of molecular chaos \cite{Zhigul}, the role of asymmetry of
functions of distribution with respect to  the transpositions of
particles \cite{koga} and many others.

So, in \cite{Khonkin} the role of large-scale  fluctuations had
been accounted for consideration of equations of two-particle
functions of distribution. The role of fluctuation is considered
in  book \cite{klimontovich2} as well. It would be desirable to
take note of book \cite{grech} which is not devoted especially to
the problem of turbulence but contains a lot of results having a
direct relation to it.

Thus, in principle, it is known how to construct  equations
corresponding to the levels of reduced  description in a general
abstract form. As a rule,  theoretical physics is content with the
qualitative investigations of  these equations. However, concrete
information extracted from them is difficult   because of their
complication. At the same time, for hydrodynamics, it would be
desirable to have the macroeffects description. In this case the
study of model equations can help. At this  approach the
Navier-Stocks equations and equations for the Maxwell's media can
be regarded as model equations. As mentioned above, the usefulness
of model equations for consideration of turbulence is confirmed by
both experimental data on the turbulent flows and their derivation
 from the Boltsman equations.  Note that, they can be obtained at the model
exponential kernels of transfer from the relations
(\ref{eq11_12}).

It is possible also to assume that the same relations with
exponential kernels will be a model in some sense for equations
with the extended set of variables in comparison with the
hydrodynamic one, describing the average flows. Characteristic
times of relaxation determined by large-scale fluctuations can be
large enough  in comparison with the time of free running of
molecules, as
 experiments demonstrate. Moreover, it is possible to make the
assumption that the equations of hydrodynamics with memory are in
some sense the case of "general position" at the modelling of
turbulence, while the Navier - Stocks system is a degenerate case,
when the time of relaxation tends to zero.

So, it is possible to make the conclusion that, at the modelling
of turbulence the equations with memory are more preferable.
Therefore, in order to make some features of turbulence on a
hydrodynamic stage more exact we should use the presented
equations or their consequences.

It turns out that the  consideration of equations with memory or
their  model  equations lead to some unexpected results and raise
problems, some of which are given in the following sections.

Also note that, some reduction procedures can be applied further
for hy\-dro\-dy\-namic equations. For example, the search of a
small number of leading parameters in the dissipative structures
theory or synergetic and phase transition theory also may be
considered as reduction in description (see the works by
I. Prigogine, H. Haken, W. Ebeling). Also remember  the new investigations
of reducing partial differential equations to low-dimensional
ordinary differential equations and their attractors
\cite{temam,lady}. The next stage of reduction consists of the
transition to the pure thermodynamic description. It should be
stressed that, in general,  the precise abstract equations for
different hierarchical levels of description are known from the
theoretical physics. Usually theoretical physicists explore these
very complex abstract equations (frequently qualitatively). But on
a hydrodynamic level it is especially interesting to search
visible macroeffects. In such a case the consideration of model
equations is especially useful. Thus, as the Navier-Stokes
equations and their hyperbolic counterpart and low-dimensional
systems of ordinary differential equations are the model
equations.

There is a great interest  in the investigation of  typical models
equations for different hierarchical levels and their typical
solutions. Note that, usually such typical solutions are the basic
elementary objects for a description of complex real phenomena.

\section { Collapses, elementary objects and instabilities}

As  was mentioned in  the previous sections, various formations
play an important role in turbulence. To some degree, the type of
structures admitted to consideration  determines possible
functional spaces. Therefore, the results presented here largely
join the results of the previous section.

Among  elementary objects,  we should first distinguish the
singularities in solutions. In other words, these are  collapses,
modes with intensification, blow-up solutions. These solutions
tend to infinity at finite time in certain points. Despite the
exoticism such solutions are often used  in the investigations of
physical processes. It turns out that they have a long history
connected with  problems of turbulence. For example,  vortical
tubes in a liquid at the pull increase  the rotation. The
dissipation of energy in turbulence takes place  not in whole
liquid, but it is concentrated in the certain localized regions
known as intermittency. Vortices can grow under the conditions of
rotations of turbulent liquids.

Naturally, the phenomena of such types are represented  in the
mathematical models of turbulence. One class of effects is
connected with the vortical motions of ideal incompressible liquid
which is a good model for the study of the developed turbulence,
when the Reynolds number  Re $\rightarrow 0$ and the influence of
viscosity is ignored. Except for pure experimental works there is
a large number of investigations  on the numerical modelling.
According to the results of modelling it has been revealed  that
under the conditions of accuracy computations with the help of
special numerical schemes, taking into account the character of
approximating methods, there is a tendency for the increase of
vorticity  (see  \cite{Krasny}). Finally, the exact analytic
solutions with vorticity collapse are found.

Singularities of initial Navier-Stokes system are a much more
difficult problem. Let us recall  the block of  ideas related to
the theory of  turbulence. This idea consists of the supposition
that the solution or its functionals are bounded only to the
finite moment of time, and then  tends to  infinity. After the
moment of collapse the solution "becomes"  finite again, next the
process repeats. Note that non-uniqueness can be observed.
Therefore, we should keep in mind the classical investigations on
the theorems of existence and uniqueness of solutions  of the
Navier-Stocks equations. Remember that  Leray proved the existence
in the three-dimensional case of a global weak solution in the
class of quadratically  summable functions under the condition
that the initial data has  bounded energy.

However, the existence of such solutions does not guarantee the
absence of singularity in the solution (i.e. the movement of a
solution to infinity). It was  also unknown, whether  such
solutions can develop from smooth initial data. The essential
results in this direction are a derivation of the strong
estimations above for the Hausdorff dimension. It turns out that
the measure of the singularities support  is equal to zero.

Recently, this problem has been developed in papers related to the
multi-valued  solutions. As a consequence, the concept of
"concentration in the solution" appeared. This means that
 sequence poorly converging in $L_2$ functions can have a
limiting function with the infinite values of solutions on the
"emaciated" set. As an example, the sequence of functions in $R^n$
\cite{majda} is presented. The work of F.Merle is indicated in
paper \cite{majda}, where the behaviour of a solution after the
moment of infinity is considered.

Let us pay attention to the work in \cite{Maystrenko} which deals
with  solutions with singularity. In this work,  for the Burgers
equations and the boundary problem, the classes of stationary
solutions having singularities were studied.

Perhaps the first solution of the Navier-Stokes system with a
collapse is considered in \cite{Krasnev}. In connection with a
collapse let us point out that investigations  were carried out
for the model equations of the hyperbolic type of the second
order.

A collapse in the system of hydrodynamics with the internal
degrees of freedom is considered numerically  by V.A.Khrisheniuk.
These results relate to strict mathematical assertions. In spite
of many open questions, the solutions with  a singularity of
vortical type were repeatedly used in different physical
investigations.  So, in \cite{Novikov68} the dynamical system
consisting of a small number of vortices  is considered and the
conditions of  chaotic motions appearance are stated. The
subsequent  investigations were developed toward the increase of a
number of the considered vortices, grates of vortices  and the
distributing of vortices  in media, when statistics of
objects-vortices  were built. We also indicate  chaos  built on
co-operation of vortices and spiral waves.

In connection with the hydro- and  gas-dynamics we  mention
another type of singularities related to the so-called gradient
catastrophes when solutions remain bounded but the derivatives on
space grow without   restriction. Thus,  there can be shock waves
in the solution. Such a situation is especially characteristic for
the compressible media. Moreover, such shock waves or solutions
formed from their combinations serve the elementary solutions in
some models of turbulence when a model is built on statistics of
such excitations (for example, turbulence at the Burgers model).

 V.E.Zaharov and co-workers dealt with collapses
considering them both numerically and analytically
\cite{zaharov1,zaharov2}. Compressed, but very capacious review is
presented in \cite{zaharov3}, where the possible applications of
collapses to hydrodynamic turbulence are announced. The role of
collapses is especially interesting in plasma as places of
dissipation (flow of energy), and also their role in Lengmyurs
turbulence, when high-frequency vibrations are concluded in a
diminishing in size cavity (to the cavity). These phenomena were
predicted in a theory, after in their confirmation the
computations  were executed, and quite recently they found
experimental confirmation \cite{Wong}. It is important, that this
process can go in retrograde on the spectrum of waves, i.e. energy
is passed in a short-wave region. It differs from many models in
which energy is passed on a spectrum in a long-wave region, where
condensation of long-wave excitations can become even similar to
the Bose - condensations in the quantum theory. In \cite{zaharov3}
the types of collapses, the volume of energy involved  and the
threshold effects had been considered.  Thus, by basic model
equations (as well as in many other works) it is the nonlinear
Shredinger equation. Because of the importance of the hypotheses
mentioned  in \cite{zaharov3} we will quote two extracts from
skies (p.470):

"For many types of turbulence, carried out in a continuous media,
the multiple development of wave collapses is characteristic in
which dissipation of turbulent energy occurs. It is not set
presently, whether classic turbulence of ideal incompressible
liquid belongs to the given type, although for this hypothesis
there are  very serious grounds...however physics of plasma is the
basic "user" of theory of wave collapses."

As our investigations have shown, the consideration of more
general models of hydrodynamics taking into account the effects of
memory, indeed give serious arguments on behalf of such a
hypothesis. Note that  the role of singularities in turbulence was
described in \cite{kulm85}. Dispersion in the system is also
important. In connection with collapses soliton turbulence was
also considered, when modulation instability was the source of
appearance of solitons as elementary objects making turbulence.

Paper \cite{Rand}, in which collapses  in a conceptual plan are
considered from a new point of view, recently appeared  (this
paper intersects partially with the cited articles). It also deals
with the solutions with the development of collapses on the basis
of the nonlinear Shredinger equation in order to adequately
describe
 abnormally large transfer in the turbulent flows. A model
is built in such a way that the solutions with collapses do not
always  grow and at large amplitudes limiting factors resulting in
the reduction of amplitude. This process repeats oneself by
spontaneous appearance and is named the homoclinic excursion. It
looks like that a model indeed passes some lines to turbulence.
However, the authors of \cite{Rand} note the question the
applicability of the Shredinger equations to the real hydrodynamic
turbulence. Let us note  that equations taking into account
memory  allow to consider similar effects.

As numerous investigations show, which are usually described in
the practically liked books on turbulence and chaos, structures
usually arise as a result of display in the systems of a different
sort of nonstability, thus the type of instability determines a
character, size and dynamics of structures. Therefore, upon
completion of this paragraph we will briefly discuss different
types of nonstability, because lately definite advancements
appeared in this direction. In addition, as is generally known
from works on dynamic chaos, it is considered that local
instability provides this phenomenon.

There is an enormous number of investigations on hydrodynamic
instability. Here, we refer  only to a small number on topics. The
first topic concerns the problems of incorrectness in physical
problems. It is known that, mostly physical problems try to
formulate so that they should be correct according to Adamar. At
the same time whole classes of ill-conditioned problems exist.
Concerning hydrodynamics ill-conditioned problems a raise  as
well. We should
 recall, for example, the problems  about explosive
instability at the Burgers equations  with the negative
coefficient of viscosity, and the problems of the boundaries of
flows (the Kelvin-Gelmgoltz instability). A non-correctness arises
 in the problems  of motion of visco-elastic  media  at the
change of their type from hyperbolical to elliptic. The Cauchy
problems  for elliptic equations  arise in  investigations of
elliptic equations  for defects. Recently, ill-conditioned
problems were considered by V.P.Maslov \cite{Maslov}. He gave the
definitions of degree of non-correctness \cite{Maslov2}.

In hydrodynamics  Maslov  considered the case of asymptotics on
viscosity  of $v\rightarrow0$ with fast-oscillating initial data,
leading to asymptotical uniqueness  and the loss of causality in
deterministic problems \cite{Maslov}. With collapses, unsteady
problems were considered, when the solution  grew due to
resonances.

Here, we briefly describe the phenomenon of waves with negative
energy in media with memory. The first such waves were considered
in the problems of electronics in the 1950's. The essence of the
phenomenon consists  in the distributed active media's definite
wave disturbances which are such characteristics at growth of
their amplitude, that simultaneously with the energy of the system
of media+wave diminishes \cite{kadomtsev2}. Formally, it is
expressed in the negative distribution of energy for these waves.
Essentially,  such effects are possible only in thermodynamically
non-equilibrium media. It is also known  that frequent
co-operation of waves of negative energy results in collapses and
that properties of such systems are determined by dispersion
correlations of linearized problems.

In conclusion we will indicate another aspect, related to
instability, turbulence and chaos. As already mentioned, for chaos
(in the mathematical sense) in the finite-dimensional  systems
instability (the Oseledets theorem) is needed. However, this
situation  is not simple for the infinite-dimensional  systems.

 In addition,
there is the separate complex problem related to the supervision
of chaos in the infinite-dimensional  systems and their
interpretation. In a manner, this task is similar with the task of
image regeneration, which, as is generally known, is improper.
There are many  examples of effects of this type and these
problems deserve a deeper study. We need  to note that in the
problems of stability  there are also intricate mathematical
problems  for example the study of stability in systems with
singular coefficients (as in the Shredinger equation with singular
potential).

\section{ Low-dimensional models for
two-dimensional generalized hydrodynamics}

Among approaches of the investigation of hydrodynamics equations
is the Galerkin method. Using this method it is easy to construct
a low-dimensional dynamical system. So we shall deal with
dynamical systems, obtained from nonlocal hydrodynamic models,
presented above.

Since generalized models can be regarded as singular perturbations
of  simpler models, it is useful to carry out the comparison of
the solutions of dynamical systems, squeezed out from different
hydrodynamics models.

In particular,  such test-systems are the well-known Lorenz system
and the low-dimensional system for plane flows, investigated by
C. Boldrighini and N. Franceschini in  1979  \cite{bold}.

\subsection{Finite-dimensional systems for 2-dimensional flows with a periodicity
condition}

 One of the models, taking into account non-locality
effects,   in the case of two spatial dimensions is the Oldroyd
fluid (the generalization of the Maxwell media (\cite{oskol, osk, mak_swir2005})
\begin{equation}\label{q21}
\begin{array}{l}
\displaystyle  \frac{{\partial \bar V}}{{\partial t}} + V_k
\frac{{\partial \bar V}}{{\partial x_k }}
 + \tau \left( {\frac{{\partial ^2 \bar V}}{{\partial t^2 }} +
 \frac{{\partial V_k }}{{\partial t}}\frac{{\partial \bar V}}{{\partial x_k }} +
 V_k \frac{{\partial ^2 \bar V}}{{\partial t\partial x_k }}} \right) - \vspace{0.25 cm}\\
 \displaystyle -\nu \Delta \bar V
 - \mu \frac{{\partial \Delta \bar V}}{{\partial t}}
   =  - \left( {1 + \tau \frac{\partial
}{{\partial t}}} \right) \mbox{grad }P + \bar F\left( {x,t} \right), \vspace{0.25 cm}\\
 \mbox{div }\bar V = 0,
 \end{array}
\end{equation}
 where $\bar V = \left\{ {V_1,\,V_2} \right\}$ are velocity components,
$\nu $ is the viscosity coefficient, $P$ is the pressure, $\bar F
= \left\{ {\bar F_1 ,\bar F_2 } \right\}$ are external forces,
$\tau $ is the relaxation time, $\mu $ is the non-locality
coefficient, $\left(  \cdot  \right)_k $ means summation over
repeated indexes.

We considered  model (\ref{q21}) on a plane region $T^2 = [0,
2\pi]\times [0, 2\pi]$ with periodical boundary conditions. The
geometry of flow and boundary conditions were imposed just like in
\cite{bold}.  In such a case we consider the flows with the
velocity component $v_z = 0$. We assume that the flow is
space-periodic in the $(x, y)$ - plane with the periods $2\pi$.
This implies that the flow and flow derivatives are periodic with
the period $2\pi$. Moreover, we also assume the mean flow averages
over  region $T^2 $ (see \cite{bold} and \cite{temam}, pp. 103
104):
$$
\int_{T^2}\hat V \,dx=0.
$$
 In our case, the solution is considered as the series on
harmonics $\exp(i\hat k \hat x)$, where $\hat x$ are coordinates
and $\hat k$ are wave vectors with integer components,
\begin{equation}\label{q23}
 \hat v\left( {x,t} \right) = \sum\limits_{k=1}^{\infty} {\gamma_k(t)\exp \left( {i k x} \right)},
\end{equation}
where $k$
is the wave vector.

Inserting (\ref{q23}) into (\ref{q21}),
after long calculations we obtained the system of equations for
coefficients $\gamma_k(t)$,
\begin{equation}
\begin{array}{l}
 \displaystyle \tau\frac{\partial^2 \gamma _k}{\partial
t^2}+\frac{\partial \gamma _k}{\partial t}= \vspace{0.25 cm}\\
\displaystyle -i \sum_{k_1+k_2+k}
\gamma_{k_1}\gamma_{k_2}(k_1+k_2)\left[1-\frac{k_1(k_1+k_2)}{|k|^2}\right]-
\vspace{0.25 cm}\\
-\nu |k^2|\gamma_k +f_1(k)\frac{k_2^2}{|k|^2}
\vspace{0.25 cm}\\
\displaystyle  -i\tau
\left[\sum_{k_1+k_2+k}\left(\gamma_{k_1}\frac{\partial\gamma_{k_2}}{\partial
t} \gamma_{k_2}\frac{\partial\gamma_{k_1}}{\partial t}
\right)(k_1 + k_2)(1 - k_1 - k_2)\right] \vspace{0.25 cm}\\
 \displaystyle +\tau\frac{\partial f_1}{\partial
t}\left(1-i\frac{k_1}{k^2}\right)-i\tau\frac{k_1}{k^2}\frac{\partial
f_2}{\partial t},
\end{array}
\end{equation}
where $f_1$, $f_2$ correspond to Fourier expansion of function
$F$.
 This infinite dimensional system is equivalent to the
original system of PDEs, yet its investigation is a very difficult
problem.

Therefore it is of common practice to "cutoff" such a system and
merely consider  a finite number of its components. Truncated
systems usually give some information about the original
equations' behaviour. Thus we must take $k$ in (\ref{q23}) from
some bounded set $L$. Let us take $L = \{ k_1 = \left( {1,1}
\right),\, k_2 = \left( {3,0} \right), \,k_3 = \left( {2, - 1}
\right),\, k_4  = \left( {1,2} \right),\, k_5  = \left( {0,1}
\right) \} $ plus opposite in sign vectors. After reduction, in
\cite{bold} the low-dimensional system was obtained $(\nu=1.0)$:
\begin{equation}\label{mak:bold}
\begin{array}{c}
    \dot \gamma_1=2\gamma_1+4\gamma_2\gamma_3+4\gamma_4\gamma_5 ,\quad  \dot
    \gamma_2=-9\gamma_2+3\gamma_1\gamma_3,\vspace{0.25 cm}\\
       \quad  \dot \gamma_3=-5\gamma_3-7\gamma_1\gamma_2+r, \vspace{0.25 cm}\\
            \dot \gamma_4=-5\gamma_4-\gamma_1\gamma_5, \quad    \dot \gamma_5=-\gamma_5-3\gamma_1\gamma_4.
\end{array}
\end{equation}

The memory effects lead to the ten-dimensional counterpart of
(\ref{mak:bold}):
\begin{equation}\label{q24}
\begin{array}{l}
\displaystyle  \frac{{dx_1 }}{{dt}} = \frac{{- x_1  - 2x_6  + 4x_7
x_8 + 4x_9 x_{10}}}{\tau } +
\\
+4\left( {x_2 x_8 + x_7 x_3 } \right) +
\\
\qquad +4\left( {x_4 x_{10}  + x_9 x_5 } \right)- 2\mu x_1 ,\vspace{3mm} \\
\vspace{3mm}  \frac{{dx_2 }}{{dt}} = \frac{{ - x_2  - 9x_7  + 3x_6
x_8 }}{\tau } + 3\left( {x_1 x_8  + x_6 x_3 } \right) - 9\mu x_2 ,
\\ \vspace{3mm}  \frac{{dx_3 }}{{dt}} = \frac{{ - x_3 - 5x_8  -
7x_6 x_7 }}{\tau } - 7\left( {x_1 x_7 + x_6 x_2 } \right) - 5\mu
x_3  + \\ +\frac{R}{\tau } , \\\vspace{3mm} \frac{{dx_4 }}{{dt}} =
\frac{{ - x_4  - 5x_9 - x_6 x_{10} }}{\tau
} - \left( {x_1 x_{10}  + x_6 x_5 } \right) - 5\mu x_1, \\
\vspace{3mm} \frac{{dx_5 }}{{dt}} = \frac{{ - x_5  - x_{10}  -
3x_6 x_9 }}{\tau } - 3\left( {x_1 x_9  + x_6 x_4 } \right) - \mu
x_5 , \\\vspace{3mm}  \frac{{dx_6 }}{{dt}} = x_1 ,\quad
\frac{{dx_7 }}{{dt}} = x_2 ,\quad \frac{{dx_8 }}{{dt}} = x_3 ,\\
\frac{{dx_9 }}{{dt}} = x_4 ,\quad \frac{{dx_{10} }}{{dt}} = x_5.
\end{array}
\end{equation}
Now we are going to describe some properties of system
(\ref{q24}). It is easily seen that stationary points for original
variables $\left( {x_6 ,...,x_{10} } \right)$ coincide with the
stationary points for 5-dimensional systems from \cite{bold}
(because in steady states for (\ref{q24}) $x_i  \equiv 0$, $i =
1,...,5$ must be fulfilled). This is because the complement terms
in (\ref{q21}) with coefficients $\tau $, $\mu $ have time
derivatives. Thus, the difference in stationary points between the
usual and generalized cases consist in stability properties. Full
investigation of ODE systems for generalized hydrodynamics is a
forthcoming problem (especially in cases of higher-dimensional
systems), so now we present only some numerical results.

Numerical calculations of the dynamical system (\ref{q24}) had been performed.
It demonstrates complicated new irregular behaviour of the solution.
This type of phase portrait ( called the butterfly) corresponds
to the Lorenz-type chaos.
Behaviour is much more complicated with memory accounting. It is characterized by the
large value of the  maximal Lyapunov exponent (which is larger
than 1) and the broad power spectrum. The projections of
trajectories have a cross-like shape. Let us note that, such forms
were found earlier in the 7-dimension system for the Navier-Stokes
equations \cite{franch}. In analogy with \cite{franch} we may
suppose that such a pattern  is created when the  eigenvalue of
the Poincare map for  the periodic orbits crosses the unit circle
in  point +1 and when  nonsteady hyperbolic-type orbits are
created.

According to the numerical integration, if $R$ is increased, the
amplitude of the limit cycle grows till stability loss, because
the period double
 bifurcation occurs. Subsequent development of the periodic regime was studied with the help
 of the Poincare bifurcation diagram. We can distinguish several period double bifurcations attached  to the  chaotic region in the diagram. It is possible to make from the conclusion  form  of the chaotic region
 that, two different types of chaotic attractors interact in the phase space of
 dynamical system (\ref{q24}).

Thus it follows from obtained results that memory
effects can cause more complicated behaviour of  solutions of the
hydrodynamical model, while  nonlocality can play the stabilizing
role (at least in considered cases).

\section { Low-dimensional model for a three - dimensional case}

Low-dimensional dynamical systems were constructed for three
dimensional flows.  Among results for such  system it is necessary
to point out two new features. Firstly, the flow with external
force demonstrates  the intermittency-II-type behaviour  under
certain conditions. Secondly, autooscillations were observed even
without the external forces.

Continuing  investigations of  model (\ref{q21}), let us take the
boundary conditions for the three-dimensional system (\ref{q21})
in the following form
\begin{equation}\label{q31}
\left. {\bar V} \right|_{\partial \Omega }  = 0,
\end{equation}
where $\Omega \in R^3 $ is a region, bounded by  surface $\partial
\Omega$, which  is the rigid boundary of the fluid volume.
Relation (\ref{q31}), known as "stick" boundary condition, means
that velocity $\vec  V$ is vanishing on the rigid boundary
(nonslip boundary conditions). Physically, it means that fluid
doesn$t$ move near  boundary $\vec V_{\partial \Omega}~=~0$.

 According to the Galerkin method the solutions of  model (\ref{q21}) are looked for as series expansions
\begin{equation}\label{q33}
V\left( {x,t} \right) = \sum\limits_k {z_k \left( t \right)\Phi _k
\left( x \right)},
\end{equation}
where $\left\{ {\Phi _k \left( x \right)} \right\}$ form the full
system of orthogonal eigenfunctions for the linear eigenvalue
problem for eigenvalues $\mu_k$ and eigenfunctions $\Phi_k$
\begin{equation}\label{q34}
\begin{array}{c}
\nu \Delta \Phi _k  =  - \mu _k \Phi _k  + \mbox{grad } p_k  ,
\quad \mbox{div } \Phi _k = 0,\, \vspace{0.25 cm}\\ \left.  {\Phi
_k } \right|_{\partial \Omega } = 0,\quad
\int\limits_\Omega  {\Phi _k \left( x \right)dx}  = 1.
\end{array}
\end{equation}
After some lengthy computations we get system for amplitudes
$z_k$, $k \in N$
\begin{equation}\label{q37}
\begin{array}{l}
\displaystyle  \tau \frac{{d^2 z_\ell  }}{{dt^2 }} +
\frac{{dz_\ell }}{{dt}} + \sum\limits_k {\sum\limits_m {c_{km\ell
} z_k z_m } } + \mu _\ell z_\ell
\vspace{0.25 cm}\\ \displaystyle + \tau \sum\limits_k {\sum\limits_m {\left\{ {z_m \frac{{dz_k }}{{dt}} + z_k \frac{{dz_m }}{{dt}} + \nu \frac{{dz_\ell  }}{{dt}}} \right\}c_{km\ell } } }  = f^\ell  , \\
 \end{array}
\end{equation}
where $\mu_\ell$, $\Phi_\ell$ are given by (\ref{q34}),
\begin{equation}\label{q38}
f^\ell   = \int\limits_\Omega  {\bar F\Phi ^\ell  dx}, \,
c_{km\ell} = \int\limits_\Omega  {\left( {\Phi ,\nabla }
\right)\Phi _m \Phi _k dx}, \ell \in N.
\end{equation}
 Values $f^\ell$ and $c_{km\ell } $
 coincide with the analogous value from \cite{bruch}. It is seen that for $\tau  = \mu  = 0$
 system (\ref{q37}) coincides  with that obtained  in \cite{lady,bruch}
for the Navier-Stokes equations. To investigate the influence of
memory and non-locality on truncated Galerkin approximations we
take the system of ODEs from \cite{bruch}:
\begin{eqnarray}\label{q40}
\frac{{dx_1 }}{{dt}} &=  - \eta _1 x_1  + Ax_2 x_3  + F_1 \nonumber
,\\ \frac{{dx_2 }}{{dt}} &=  - \eta _2 x_2  + Bx_1 x_3  + F_2,\nonumber \\
\frac{{dx_3 }}{{dt}} &=  - \eta _3 x_3  + Cx_1 x_2  + F_3 ,
\end{eqnarray}
 where $A$ , $B$ , $C$ , $F_i $ are
constant and an additional relation is fulfilled $A + B + C = 0$.
The last condition is derived in  \cite{bruch} from consideration
of non-viscous fluid kinetic energy with nullifying mass forces
($\nu = 0$ , $\bar F = 0$). Let us fix parameters
\begin{equation}\label{q39}
\begin{array}{c}
A = 1, B = - 2, C = 1,\vspace{0.25 cm}\\
F_1 = - \frac{{ - 5 + \sqrt {76} }}{2}, F_2 = - 20 - \sqrt {76},
F_3 = \frac{{34 + 5\sqrt {76} }}{4}, \vspace{0.25 cm}
\\ \eta _i = \nu, \quad i = 1,2,3.
\end{array}
\end{equation} for illustration.
Then the low-dimensional system in this case is
\begin{multline}\label{q311}
 \frac{{dx_1 }}{{dt}} = \frac{{ - x_1  - x_5 x_6  - \nu x_4  +
F_1 }}{\tau } - \left( {x_3 x_5  + x_6 x_2 } \right) -\\- \nu
 \rm{sgn} \left| \mu \right|x_1, \vspace{3mm}\\ \frac{{dx_2
}}{{dt}} = \frac{{ - x_2  + 2x_4 x_6 - \nu x_5  + F_2 }}{\tau } +
2\left( {x_1 x_6  + x_3 x_4 } \right) +\\+ \nu \rm{sgn} \left| \mu
\right|x_2,  \\ \frac{{dx_3 }}{{dt}} = \frac{{ - x_3  - x_4 x_5 -
\nu x_6  + F_3 }}{\tau } - \left( {x_2 x_3  + x_1 x_5 } \right) -
\\-\nu  \rm{sgn} \left| \mu \right|x_3 , \\ \frac{{dx_4
}}{{dt}} = x_1, \frac{{dx_5 }}{{dt}} = x_2 , \frac{{dx_6 }}{{dt}}
= x_3.
\end{multline}

 It was shown in \cite{bruch} that in the case of Navier-Stokes system
there is a Hopf bifurcation, when $\nu  = 0$. Value $\nu _{cr}  =
2$ separates steady state and the region with periodic orbits.
There is only a stable equlibrium for $\bar F = 0$. Now, let us
present the results of the numerical integration of system
(\ref{q311}) performed with the help of the Runge-Kutta method.
It is seen that growth of
amplitudes is observed. Then we have decaying oscillations and
almost constant solution on time length about five time units.
However, near time 9.0 there is a strong burst of the solution.
After  time of 12.0 the burst decays. Then after time 14.0 the
burst takes place again. This is similar with type-II
intermittency.

Another interesting result is an oscillating regime without mass forces. These oscillations take place in
nonzero initial values. This case can be interpreted as an
evolution of flow without external influence. The presented
behaviour illustrates one of the possible ways of explanation of
turbulence in  flows without restoring to negative viscosity. Our
computations also confirm the stabilyzing role of nonlocality.

Let us summarize the numerical and analytical results of the
investigations of systems (\ref{q40}), (\ref{q311}),
(\ref{mak:bold}) and (\ref{q24}).

 For the six-dimensional system
(\ref{q311}) we found the emergence of periodical solutions. Such
solutions were also found  without external forces $(F = 0)$ with
nonzero initial conditions. We also found, in some cases,
phenomena similar to "intermittency" (bursts in solutions). The
ten-dimensional system (\ref{q24}) was investigated for some
parameters. We changed the values of relaxation time and the
initial conditions. In the case with no memory $(\tau = 0)$ the
projections of the phase portrait on two-dimensional planes have a
"butterfly" type of attractor similar to the diagrams  for the
Lorenz chaotic attractor. In the case with $\tau \neq  0$, there
is complex behaviour of a new type. Visually, the trajectory
densely fills
 a bounded volume (named "container"). Trajectory
projections on the planes have a broken form in many points.
Locally, the projection of phase portrait looks like the  ball of
thread or "patience". Visually, the behaviour is similar to
two-dimensional mappings with homoclinic tangency and
quasiattractors which was described in \cite{gonchenko}. Similar
pictures were found in systems with the so-called "fat-fractals"
with larger fractal dimension than the Cantor-type set. In many
cases the results of numerical calculations look like  the
projections of  motion on the torus. The first Liapunov
characteristic exponent stayed positive for a long time (but with
diminishing value). Numerical investigation of (\ref{q24}) with
different values of $R$ displayed the still excitation of complex
behaviour. Standard one-dimensional bifurcation diagram is
entirely different from the usual period doubling scenario of
transition to chaos. We also made some numerical investigations of
bifurcation for the 10-dimensional system. We found that in case
of vanishing external forces $(R = 0)$ there was a unique
stationary point with zero co-ordinates. The Jacobian of the right
part of (\ref{q24}) had a pair of pure imaginary eigenvalues. Some
further bifurcation was received by increasing the number of
stationary points to ten with increasing the values of $R$. At
this processes the pairs of complex conjugate eigenvalues cross
the imaginary axis from left to right. Further investigations of
the systems above look very promising in the case $\tau \ll 1$.
This is the singular perturbation of the usual systems of ODEs
with chaos. The evaluation of bounds for the attractor dimension
in the case $\tau \rightarrow 0$ is interesting (especially in the
limit $N$ or/and $R$ tending to infinity). Note that (\ref{q37})
consists of ODEs of the second order in time. Hence, such a
system recalls the collection of oscillators. Each $z_k$
corresponds to a wave number of harmonics $k$. So, we may
anticipate the properties like transmission of energy on the
spectrum of harmonics, the existence of harmonics clusters,
resonances and so on. Also, some ideas of memory effects in
turbulence may be reconsidered. There are many space and time
scales in fluid flows. This implies the existence of many types of
chaotic behaviour in fluids. The account of memory leads
presumably to a new type of chaos similar to the chaos in media
constructed from oscillators. Let us also note that, the above
complex behaviour may serve as a prototype of a new possible type
of chaos in media with finite speed of propagation and with gauge
symmetry.

\section{Applications of model equations to physical processes}

Proposed model are rather new and are useful objects for further
mathematical investigations. But just now some interpretations of
the solutions may be proposed for real physical processes. Some
applications to heat and mass transfer processes had been proposed
and tested experimentally earlier: hyperbolic heat conduction
equation, ignition theory by heat explosion, heat and mass
transfer in turbulent and heterogeneous media (see
\cite{mak_swir2005}). New peculiarities of solutions founded in
new model equations allows considering some aspects of very
important recent physical and technical problems where strong non
� equilibrium properties are essential. Wide range of applications
supplies the processes in astrophysics. The first subclass of
problems is the description of the phenomena in the sparse plasma,
including also MHD. The second relevant field of application is
the investigation of the processes in the near-Earth space where a
great number of different structures have been found
experimentally. The third field of the investigation is the
processes at the Sun, especially forming of hot and cool spots,
arcs, bursts, cellular lattices on the surface etc. Many theories
have been proposed before but the problems are still open. The
most interesting consequences from proposed modal equations for
such problems considerations are presumable fragmentation of flow
with diminishing of scales and increasing of vorticites an!
 d of solutions amplitudes for fast flows.
Other important problem is the building of thermonuclear power plants. One of the main problems is such processes and apparatus control and suppression of wide spectrum of possible instabilities. Usually classical MHD equations are used. But proposed considerations lead to the conclusion of new more accurate equations needs far from equilibrium in such processes.
Finally proposed model equations with memory and non-locality effects can be used for extending the classical synergetic problems to processes much far from equilibrium the in usual synergetic. Note for example that in classical synergetic classical Kuramoto-Sivashinsky equation is used for considering many kinds of instabilities. So, for the case of far from equilibrium processes the new equations for instabilities investigation may be proposed. Such equations may be extending of Kuramoto-Sivashinsky equation but with accounting memory and non-locality effects. For example one of such generalized equations (for one space dimension case) for structures formation has the form:

$$
 \tau \frac{\partial ^2 \ u}{\partial t^2 }+  \frac{\partial \
u}{\partial t} +  u\frac{\partial \ u} {\partial x} +  \nu
\frac{\partial ^2 \ u}{\partial x^2 } + \alpha \frac{\partial ^3 \
u}{\partial x^3 } + \beta
               \frac{\partial ^4 \ u}{\partial x^4 } = 0,
$$
where    $\alpha,     \beta,  \mu,  \tau,   \nu$       �
parameters. Note that the ordinary differential equation for the
form of �travelling wave� solution coincides with the same in
classical case $(\tau=0)$ but the stability conditions for such
solutions are different from the classical case.

\section { Conclusions}

Thus, in this paper we propose to extend the studies of
memory effects on some distributed systems which have
classical description by partial differential and ordinary equations.
This allows
posing new mathematical problems on investigation of these new
objects and their solutions. The first studies allow to find new
properties and to propose a lot of new research problems.

\end{document}